\documentclass[a4paper,11pt]{article}
\usepackage{jheppub} 
\usepackage{color}
\usepackage{subcaption}
\usepackage[normalem]{ulem}

\title{ Extremal curves in single-trace $T\bar{T}$-holography}

\author{Soumangsu Chakraborty, Madhur Mehta, Gela Patashuri}
\affiliation{Department of Physics,
Center for Cosmology and AstroParticle Physics (CCAPP)\\
The Ohio State University,
191 W Woodruff Ave, Columbus, OH 43210, USA}

\emailAdd{soumangsuchakraborty@gmail.com,mehta.493@osu.edu, patashuri.1@osu.edu}

\abstract{In this paper, we continue the study of single-trace $T\bar{T}$-holography where the boundary field theory can be realized as a CFT$_2$ deformed by a single-trace irrelevant operator of dimension $(2,2)$ and dual spacetime geometry is $AdS_3$ smoothly glued to flat spacetime with a linear dilaton near the boundary.  In this non-AdS holographic framework, we propose that the length of real extremal curves connecting the two boundaries of an eternal black hole at fixed boundary time captures the time-evolved entanglement entropy of an entangled, quenched boundary system. At late times, we find two analytic extremal solutions in the complexified geometry, which become real in complementary temperature regimes. Focusing only on the real solutions leads to a non-analyticity at a critical temperature $T_c$, which we interpret as a second-order phase transition separating a local (CFT$_2$) phase from a non-local (Little String Theory) phase.}

\begin{document}
\maketitle
\flushbottom

\section{Introduction}\label{sec:intro}
Single-trace $T\bar{T}$-holography \cite{Giveon:2017nie} arises from the standard $AdS_3/CFT_2$ correspondence by deforming the spacetime CFT$_2$ by  a specific dimension $(2,2)$ single-trace quasi-primary operator  $D(x,\bar{x})$, constructed in \cite{Kutasov:1999xu}. Although the deformation on the boundary field theory is irrelevant, it remains under control, as it induces on the worldsheet string theory ($SL(2,R)$ WZW model at level $k$) in $AdS_3$ in the presence of pure NS-NS flux a truly marginal deformation \cite{Giveon:2017nie,Chakraborty:2024mls}. Such a marginal deformation of the worldsheet theory takes the Abelian Thirring form, which is exactly solvable.\footnote{This deformation can also be realized as a TsT transformation of the $SL(2,R)$ WZW model on the worldsheet \cite{Apolo:2019zai}.} The resulting spacetime theory, often termed as single-trace $T\bar{T}$ deformed CFT$_2$, flows from (non-local) Little String Theory (LST) at short distances to a local CFT$_2$ in the IR. 

The operator $D(x,\bar{x})$ is single-trace and shares many properties with the standard $T\bar{T}$ operator, which is double-trace.\footnote{Hence, the term ``single-trace $T\bar{T}$ operator'' is often used to denote the operator $D(x,\bar{x})$.} 
It is a massive mode of the graviton--dilaton sector of string theory in $AdS_3$ and can be added to the worldsheet Lagrangian to deform the $AdS_3$ asymptote to flat spacetime with a linear dilaton, while keeping the $AdS_3$ region in the core intact. 
This allows us to systematically study {\it holography in an asymptotically non-AdS spacetime}, which is smoothly connected to the standard $AdS_3/CFT_2$ holography in the undeformed limit. 
By contrast, the standard $T\bar{T}$ operator, as mentioned above, is double-trace. 
In the semiclassical approximation, its effect on the gravity side (after deformation) is to preserve the $AdS_3$ asymptote but change the boundary conditions for the bulk fields from Dirichlet to mixed \cite{Guica:2019nzm}. 
See \cite{Jiang:2019epa,He:2025ppz} for a review of $T\bar{T}$-deformed CFT$_2$, its extensions, and their holographic implications.

In this paper, we investigate extremal curves in single-trace $T\bar{T}$-holography. One of the most important motivations for studying extremal curves in geometries that interpolate between $AdS_3$ in the IR and linear dilaton spacetime in the UV stems from the fact that the Ryu-Takayanagi \footnote{It can be easily shown that the Ryu-Takayanagi curve also satisfies the strong subadditivity inequality of entanglement entropy in spacetime that interpolate between $AdS_3$ in the IR and linear dilaton spacetime in the UV.} proposal has been argued to accurately capture entanglement entropy of a bipartite system in the boundary field theory (LST) \cite{Chakraborty:2018kpr,Asrat:2020uib,Asrat:2019end,Chakraborty:2020udr}. 
Motivated by the proposal of \cite{Hartman:2013qma}, which applies strictly to eternal black holes in AdS, we suggest that the extremal curve connecting the two boundaries of an eternal black hole in single-trace $T\bar{T}$-holography at a fixed boundary time serves as a natural candidate for capturing the time-dependent entanglement entropy of a quenched state of the entangled boundary system ({\it i.e.}, two entangled copies of LST). We further support this proposal through a couple of consistency checks.

We consider a two-sided eternal black hole solution in single-trace $T\bar{T}$-holography, obtained in the decoupling limit of a certain configuration of NS5 branes and F1 strings in type II string theory, and consider an extremal curve connecting the two boundaries at a given instant of the boundary time. 
We show that there are two such extremal solutions, each of which is analytic as a function of temperature in the maximally extended complexified geometry. 
When restricted to the real geometry, at late boundary times, the extremal curves are real in complementary temperature domains. 
We show that the curves, when restricted to the real plane, exhibit non-analyticity at some critical temperature $T = T_c$.  
Below the critical temperature, one of the extremal solutions becomes complex, while above it, the other solution ceases to be real. 
We find that the lengths of both extremal curves are identical at $T = T_c$ at late times. 
Our proposal—that the length of the real extremal curve computes the time-dependent entanglement entropy of the quenched boundary system implies that, as the temperature crosses the critical temperature, the extremal solution must jump from one to the other. 
This jump induces a discontinuity in the second derivative of the entanglement entropy at late times, which we interpret as a second-order phase transition separating the local phase from the non-local phase.

The paper is organized as follows. 
In Section \ref{sect:geometry}, we introduce the type II SUGRA background that corresponds to an eternal black hole in single-trace $T\bar{T}$-holography and discuss its thermodynamic properties. 
In  Section \ref{sec3.1}, we review the computation of extremal curves connecting the two boundaries of a non-spinning BTZ black hole, and in Section  \ref{sec3.2} we extend the analysis of  Section  \ref{sec3.1} to an eternal black hole in single-trace $T\bar{T}$-holography and discuss its physical interpretation from the perspective of LST. 
Finally, in Section \ref{diss}, we summarize our findings and discuss potential directions for future research.

\section{The SUGRA background} \label{sect:geometry}

We begin with type II string theory on $R_t \times S_x^1 \times T^4 \times R^4$, with $k$-NS5 branes wrapping $S_x^1 \times T^4$ and $p$-F1 strings wrapping the circle $S_x^1$ with radius $R$. Upon back reaction, the non-extremal configuration is described by the black brane background (in string frame) \cite{Maldacena:1996ky,Hyun:1997jv}\footnote{The three form field strength $H$ will not feature in our calculation. We will therefore omit it in the rest of the discussion.}
\begin{equation}
\begin{aligned}
&d s^2 =\frac{1}{f_1}\left(-f d t^2+d x^2\right)+f_5\left(\frac{1}{f} d r^2+r^2d \Omega_3^2\right)+\sum_{i=1}^4 d x_i^2~, \\
&e^{2 \Phi}=g^2 \frac{f_5}{f_1}~,\\
&H=d x \wedge d t \wedge d\left(\frac{r_0^2 \sinh 2 \alpha_1}{2 f_1 r^2}\right)+r_0^2 \sinh 2 \alpha_5 d \Omega_3~,
\end{aligned}
\end{equation}
where $x_i$ represent the coordinates on $T^4$, and  $\left(r, \Omega_3\right)$ are the spherical coordinates in  the $R^4$ transverse to the fivebranes. The geometry has an outer horizon at $r=r_0$ and an inner horizon at $r=0$.
The constant $g=e^{\Phi(r \rightarrow \infty)}$ is the asymptotic string coupling far from the fivebranes,   related to the ten-dimensional Newton constant in flat space
\begin{equation}
    G_N^{(10)}=8 \pi^6 g^2 l_s^8~,
\end{equation}
where $l_s=\sqrt{\alpha'}$ is the fundamental string length. The harmonic functions of the fivebranes and F1 strings and the blackening factor $f$ are given respectively by
\begin{equation}
    f_{5,1}=1+\frac{r_{5,1}^2}{r^2}~,\quad  f=1-\frac{r_0^2}{r^2}~,
\end{equation}
where
\begin{equation}
    r_{1,5}^2=r_0^2 \sinh ^2 \alpha_{1,5}~,\quad \sinh 2 \alpha_1= \frac{2 (2\pi)^4 p g^2 l_s^6}{V _{T^4}r_0^2}~, \quad \sinh 2 \alpha_5= \frac{2 l_s^2 k}{r_0^2}~,
\end{equation}
with $V_{T^4}=(2\pi)^4v l_s^4$ being the (dimensionful) volume of $T^4$. 
\subsection{Decoupling limit}

As is standard in the Little String Theory (LST) holography \cite{Giveon:1998ns}, one can decouple the theory of the fivebranes by sending $g\to 0$, with $r,r_0, r_1$ scaling like $g$. This can be obtained by the rescaling
 \begin{equation}
    r \rightarrow g r~,\quad r_0 \rightarrow g r_0~,\quad r_1 \rightarrow g r_1~,
\end{equation} 
followed by the limit $g \to 0$ while keeping the rescaled parameters $r, r_0$, $r_1$ fixed \cite{Chakraborty:2020swe,Chakraborty:2023mzc,Chakraborty:2023zdd}. In this limit, $\alpha_5$ becomes infinite, and $r_5$ approaches the value 
\begin{equation}\label{r5}
    r_5 \simeq \sqrt{k}l_s~.
\end{equation}
Note that the dynamics of the NS5 branes decouple from the bulk without taking $l_s$ to zero. This has an intriguing implication from the perspective of the boundary field theory; the boundary field theory is a strongly coupled, non-local, non-gravitational theory of strings popularly referred to in the literature as Little String Theory (LST).

In this limit, the radius of the $S^3$ transverse to the fivebranes stabilizes to $r_5$ \eqref{r5}. The resulting  background factorizes into the  an $S^3$ of radius $r_5$ and $T^4$, and a non-compact $(2+1)$-dimensional spacetime corresponds to the background fields 
\begin{equation}
\begin{aligned}
& d s^2=\frac{1}{f_1}\left(-f d t^2+d x^2\right)+\frac{r_5^2}{r^2 f} d r^2 + k \alpha^{\prime} d \Omega_3^2+d s_{T^4}^2~, 
\qquad e^{2 \Phi}=\frac{r_5^2}{r^2 f_1} ~.  
\end{aligned}
\label{eq:DecoupMetr1}
\end{equation}
The decoupled geometry \eqref{eq:DecoupMetr1}  has a transition scale $r_1$ which will play a very important role in our discussion. Without loss of generality, we can set this scale to 1. However, as it turns out, it's more convenient to introduce a new dimensionless radial coordinate
\begin{equation}
    U = \frac{r}{r_1}~,\qquad U_T = \frac{r_0}{r_1}~,
\end{equation}
which is equivalent to setting $r_1=1$.
With this choice of coordinates (removing the $S^3$ and the $T^4$ part)
the background (\ref{eq:DecoupMetr1}) takes the form 
\begin{equation}
\begin{aligned}
    \label{eq:DecoupMetr2}
    &d s^2=\frac{1}{f_1}\left(-f d t^2+d x^2\right)+\frac{k \alpha^{\prime}}{f} \frac{d U^2}{U^2}~,\\
    &e^{-2(\Phi-\Phi_0)}= f_1 U^2~ , \quad e^{2\Phi_0}= \frac{k\alpha'}{r_1^2}~,\\ 
    &f=1-\frac{U_T^2}{U^2}~, \quad f_1=1+\frac{1}{U^2}~.
\end{aligned}    
\end{equation}
In the new coordinates, the outer horizon is at $U=U_T$. It is easy to check that the geometry \eqref{eq:DecoupMetr2} asymptotes to flat spacetime with a linear dilaton. In the discussion that follows, we will refer to this background as $\mathcal{M}_3^T$ and its extremal limit, namely $U_T\to 0$, as $\mathcal{M}_3$.

\begin{figure}[t]
    \centering
    \begin{subfigure}[b]{0.55\textwidth}
        \includegraphics[width=\textwidth]{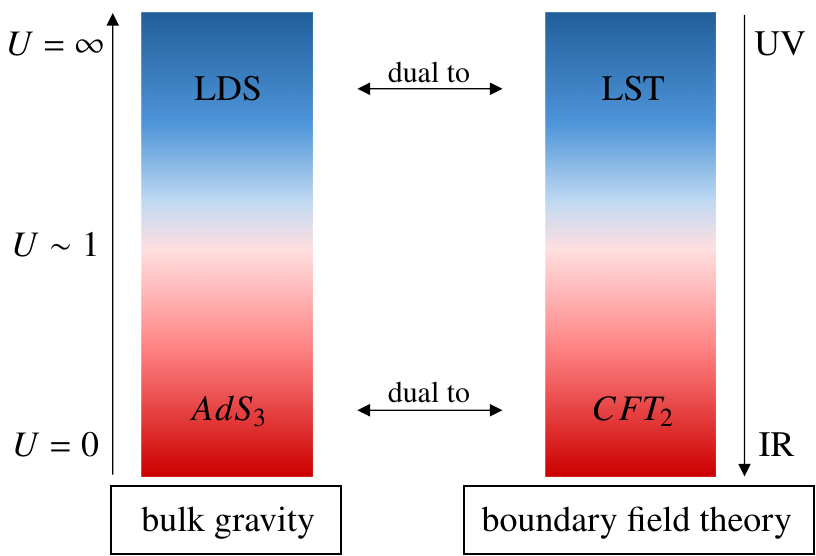}
        \caption{}
        \label{fig:1a}
    \end{subfigure}
    \hfill
    \begin{subfigure}[b]{0.395\textwidth}
        \includegraphics[width=\textwidth]{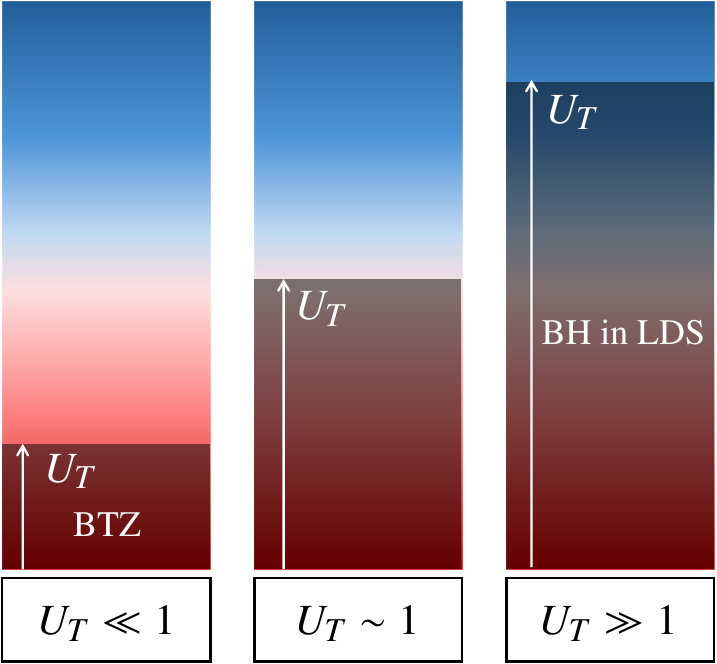}
        \caption{}
        \label{fig:1b}
    \end{subfigure}
    \caption{(a) 
    RG flows from both the bulk gravity side and the boundary field theory side. (b) Schematic representation of the black hole \eqref{eq:BTZdecoupled} that interpolates between BTZ at low temperature and a black hole in linear dilaton spacetime (LDS) at high temperature.} 
    \label{fig:sidebyside}
\end{figure}

One can further consider the decoupled theory of the fundamental strings by restricting to $U_T< U\ll1$.\footnote{This is equivalent to taking $l_s/R\to 0$.} In that case, the geometry becomes that of non-rotating BTZ.
\begin{equation}
\begin{aligned}
    &d s^2=\frac{1}{f_1}\left(-f d t^2+d x^2\right)+\frac{k \alpha^{\prime}}{f} \frac{d U^2}{U^2}~, \ \ \ \ \ \
    e^{2\Phi}= \frac{kv}{p}~ ,\\ 
    &f=1-\frac{U_T^2}{U^2}~, \quad f_1=\frac{1}{U^2}~.
\end{aligned}    
    \label{eq:BTZdecoupled}
\end{equation}
A few comments are in order
\begin{itemize}
    \item In the extremal limit ({\it i.e.}, $U_T = 0$),  the geometry (\ref{eq:DecoupMetr2}) interpolates between massless  BTZ (with $AdS$ radius $\ell=\sqrt{k}l_s$) in the IR ({\it{i.e.},} $0<U\ll 1$) and the linear dilaton spacetime in the UV ({\it{i.e.},} $U \gg 1$).

    \item From the dual boundary theory perspective, this can be visualized as an RG flow from LST at short distances to a local CFT$_2$ at long distances. Close to the IR fixed point, one can think of this flow being triggered by a dimension $(2,2)$ irrelevant operator, often termed in the literature as the single-trace $T\bar{T}$ operator. 
    See Fig. \ref{fig:1a} for a pictorial demonstration of this RG flow \cite{Giveon:2017nie,Chakraborty:2024mls}.

    \item When $U_T>0$, the geometry \eqref{eq:DecoupMetr2} interpolates between a BTZ black hole in the IR  to the black hole in linear dilaton spacetime in the UV (see Fig. \ref{fig:1b}) shown in Fig. \ref{fig:1a}.
\end{itemize}

\begin{figure}[t]
    \centering
    \includegraphics[width=0.7\textwidth]{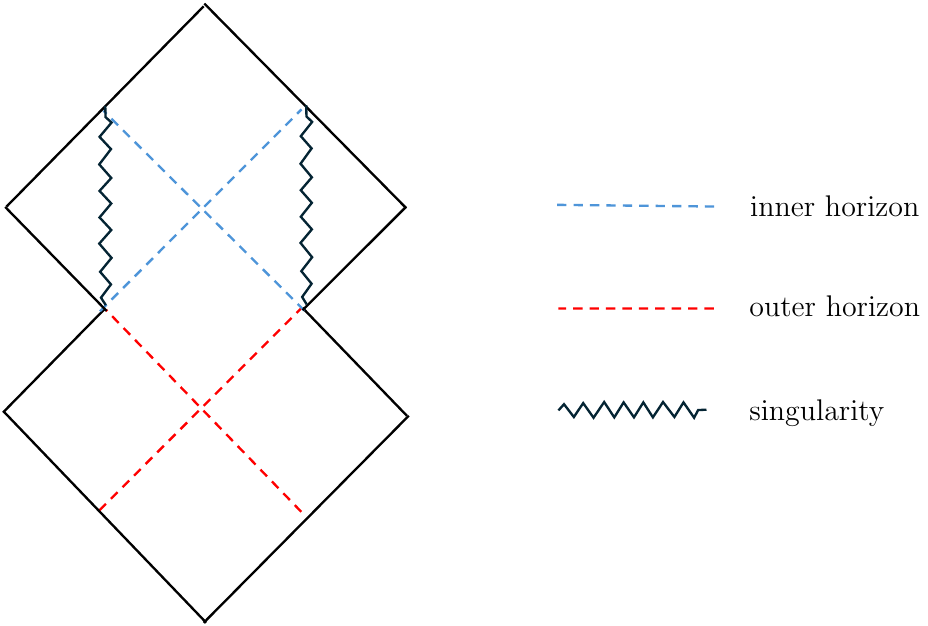}
    \caption{Penrose diagrams of the eternal  black hole $\mathcal{M}^T_3$.} 
    \label{fig:PenroseM3}
\end{figure}

For convenience, we introduce a change of coordinate
\begin{equation}
    \cosh \rho= \frac{U}{U_T}~.
\end{equation}
In this new coordinate, the background \eqref{eq:DecoupMetr2} takes the following form: 
\begin{equation}
\begin{aligned}
    \label{eq:RELEVMetrM3T}
    &d s^2_{\mathcal{M}}=-  g(\rho)^2d t^2+h(\rho)^2 d x^2+k \alpha^{\prime} d \rho^2~,\quad e^{-2(\Phi-\Phi_0)} = j(\rho)~,\\
    &g(\rho)^2 = \frac{U_T^2 \sinh ^2 \rho}{j(\rho)}~,\quad h(\rho)^2 = \frac{U_T^2 \cosh ^2 \rho}{j(\rho)}~, \quad j(\rho) = 1+U_T^2 \cosh ^2 \rho~.
    & 
\end{aligned}    
\end{equation}
Its Penrose diagram is illustrated in  Fig. \ref{fig:PenroseM3} \cite{Chakraborty:2020fpt}, depicting the causal structure of linear dilaton spacetime.  

The coordinates $\rho>0, t\in R$ cover the right wedge of the Penrose diagram \ref{fig:PenroseM3}. To describe the interior of the black hole (top wedge), one needs to analytically continue 
\begin{equation}\label{analytic}
    \rho\to i\kappa~,\ \ \ \ \ \  t\to t_I-\frac{i}{4T}~,
\end{equation}
 where $T$ is the temperature of the black hole \eqref{eq:RELEVMetrM3T}.

\subsection{Thermodynamics}

Thermodynamics of the black hole \eqref{eq:DecoupMetr2} has been extensively studied \cite{Apolo:2019zai,Chakraborty:2020swe,Chakraborty:2023mzc,Chakraborty:2023zdd}.
The black hole temperature, obtained by going to the Euclidean continuation of the geometry \eqref{eq:DecoupMetr2} and demanding smoothness at the Euclidean horizon, is given by
    \begin{equation}\label{temp}
    T=T_H \sqrt{\frac{U_T^2}{1+U_T^2}}~,
\end{equation}
where $T_H$ is the Hagedorn temperature of LST
\begin{equation}\label{th}
    T_H = \frac{1}{2\pi\sqrt{k\alpha'}}~.
\end{equation}
As expected, the temperature \eqref{temp} approaches the Hagedorn temperature, $T_H$, of the fivebranes when $U_T\gg 1$. This is a signature behavior of black holes in a linear dilaton background. On the other hand, when $U_T \ll 1$, the temperature approaches that of a BTZ black hole
\begin{equation}
    T_{BTZ}=\frac{U_T}{2\pi\sqrt{k\alpha'}}~.
\end{equation}

The Bekenstein-Hawking entropy of the black hole \eqref{eq:DecoupMetr2} is given by \cite{Chakraborty:2020swe,Chakraborty:2023mzc,Chakraborty:2023zdd}
\begin{equation}
    S = \sqrt{\frac{E^2}{T_H^2}+\frac{4\pi^2 c}{3}  E R}~,
    \label{eq:entropyInit}
\end{equation}
where $E$ is the ADM energy of the black hole above extremality and $c = 6\,k\,p$ is the central charge of the CFT$_2$ in the IR. For large energies, $RE\gg1$, the entropy grows linearly in $E$, with a coefficient given by the inverse Hagedorn temperature of LST. When $RE\ll 1$, the entropy formula \eqref{eq:entropyInit} is very well approximated by the Cardy formula, which captures the entropy of a BTZ black hole. 

Using black hole thermodynamics
\begin{equation}
    \frac{1}{T} = \frac{\partial S}{\partial E} = \frac{\frac{E}{T_H^2}+\frac{2\pi^2c}{3}R}{\sqrt{\frac{E^2}{T_H^2}+\frac{4\pi^2c}{3} E R}}~,
    \label{eq:invTemp}
\end{equation}
one can easily express the thermodynamic energy as a function of temperature
\begin{equation}
    E = \frac{2\pi^2c}{3} \left(\frac{T_H}{\sqrt{T_H^2-T^2}}-1\right)RT_H^2~.
    \label{eq:BHenergy}
\end{equation}
Substituting \eqref{eq:BHenergy} in \eqref{eq:entropyInit}, one can express the Bekenstein-Hawking entropy as a function of temperature
\begin{equation}\label{sbek}
    S = \frac{2\pi^2c}{3} \frac{ RT }{ \sqrt{1 - \frac{T^2}{T_H^2}}}~ .
\end{equation}
As usual, as $T\to T_H$ the system exhibits a Hagedorn density of states, whereas for $T\ll T_H$ it reproduces the entropy of a BTZ black hole
\begin{equation}\label{sbtz}
    S^{BTZ}  = \frac{2\pi^2c}{3} R T~. 
\end{equation}
One can think of the parameter $T/T_H$ to be the interpolating parameter that interpolates between the Cardy regime ({\it i.e.}, $T/T_H\ll1$) and the Hagedorn regime ({\it i.e.}, $T/T_H \to 1$).

The pressure of a fluid whose equation of state in thermal equilibrium is given by \eqref{eq:BHenergy} takes the form
\begin{equation}\label{press}
    P \sim -\frac{\partial F}{\partial R} = \frac{ \frac{2\pi^2c}{3}E}{ \frac{E}{T_H^2}+ \frac{2\pi^2c}{3}R}~,
\end{equation}
where $F=E-ST$ is the free energy of the system. Differentiating the pressure with respect to the thermodynamic energy gives the square of the velocity of sound in the medium
\begin{equation}
    v_s^2 \sim \frac{\partial P}{\partial E} = 1-\frac{T^2}{T_H^2}~.
    \label{eq:velOFwave}
\end{equation}
We conclude this section with the following comments.
\begin{itemize}
    \item The velocity of sound goes to zero as $T\to T_H$. This is a typical behavior of the propagation of mechanical vibrations in LST \cite{Parnachev:2005hh,Goykhman:2013oja,Chakraborty:2022dgm}. 

    \item The vanishing of the velocity of sound in the deep UV is a semi-classical result; worldsheet genus corrections give non-zero corrections \cite{Parnachev:2005hh}. Such higher genus corrections are usually heard. Realizing LST as an irrelevant deformation of a CFT$_2$ allows us to calculate the velocity of sound, in the semiclassical approximation, at all points along the RG flow. 

    \item The velocity of sound goes to unity at low temperature in exact agreement with our CFT$_2$ intuition.

    \item The thermal entropy, $S(E)$, in double-trace $T\bar{T}$ deformed CFT$_2$ is identical to \eqref{eq:entropyInit} (at high energies) \cite{Datta:2018thy,Chakraborty:2020xyz}, even though the two theories; the one studied here and double-trace $T\bar{T}$ deformed CFT$_2$ are different. Since the analysis leading up to the velocity of sound only takes as an input $S(E)$ \eqref{eq:entropyInit}, the velocity of sound in double-trace $T\bar{T}$ deformed CFT$_2$ is identical to \eqref{eq:velOFwave}.   

\end{itemize}

In the next section, we will reproduce certain aspects of thermodynamics from an independent computation of extremal curves in the background \eqref{eq:DecoupMetr2}.

\section{Late time behavior of entanglement entropy }

In this section, we aim to study thermalization processes in the full boundary theory dual to string theory in the background \eqref{eq:DecoupMetr2}. 
Thermalization is a fundamentally important phenomenon in non-equilibrium physics. 
A simple model for thermalization is provided by a quantum quench, where the Hamiltonian of the system is suddenly changed, leading to the time evolution of the ground state of the new Hamiltonian. 
However, studying quantum quenched systems is considerably challenging if one relies solely on coarse-grained physical quantities such as correlation functions. 
For example, it is difficult to distinguish between pure state of a quantum quench and a thermal mixed state \cite{RangamaniTakayanagi2017}. 
This distinction, however, can be probed using entanglement entropy, which serves as a more powerful tool that provides deeper insights into the system.

\subsection{Extremal curves in the BTZ spacetime---a review}\label{sec3.1}

A quantum quenched one-dimensional system has already been studied in the holographic framework
 \cite{Hartman:2013qma}, where the time evolution of the entanglement entropy—previously derived from a CFT$_2$ computation \cite{Calabrese:2005in}—was reproduced by probing the interior of an eternal BTZ black hole. The main tool used was the area of an extremal curve $\gamma$ in AdS,  
anchored at the boundary of the entangling region (a spacelike surface) residing on the boundary of AdS. According to the proposal in \cite{Ryu:2006bv, Hubeny:2007xt},  such curves are related to the entanglement entropy of the boundary field theory via $S_e = \frac{Area(\gamma)}{4G_N}$.
The goal of this subsection is to briefly review this computation of \cite{Hartman:2013qma} for the specific case of a non-spinning BTZ black hole.

To begin with, let us  consider the maximally extended eternal non-spinning BTZ black hole \eqref{eq:BTZdecoupled}, \footnote{Note that, unlike \cite{Hartman:2013qma}, we do not fix the temperature, since in later analysis --- when we consider the $\mathcal{M}_3^T$ geometry --- temperature becomes an important parameter that influences the physics we are interested in.}:
\begin{equation}
\begin{aligned}
    d s^2_{\text{BTZ}}=-M  \sinh ^2 \rho d t^2+M\cosh ^2 \rho d x^2+\ell^2 d \rho^2~,
\label{eq:BTZspacetime}
\end{aligned}    
\end{equation}
where a dimensionless mass parameter, $M$, and the radius of $AdS_3$, $\ell$, have been introduced; 
\begin{equation}
    M = U_T^2 = \left(\frac{T_{BTZ}}{T_H}\right)^2,\qquad \ell = \sqrt{k\alpha'}~.
\end{equation}
The metric \eqref{eq:BTZspacetime} can be obtained by the standard BTZ decoupling limit (see {\it e.g.}, the discussion around \eqref{eq:BTZdecoupled} and \eqref{eq:RELEVMetrM3T}): $U^2 = U_T^2\cosh^2\rho \ll1$ in (\ref{eq:RELEVMetrM3T}). Note that, in this limit, the dilaton becomes a constant as expected. In the $\rho$ coordinate, $\rho=0$ is the horizon and $\rho=i\pi/2$ is the BTZ singularity.

\begin{figure}[t]
    \centering
    \begin{subfigure}[b]{0.35\textwidth}
        \includegraphics[width=\textwidth]{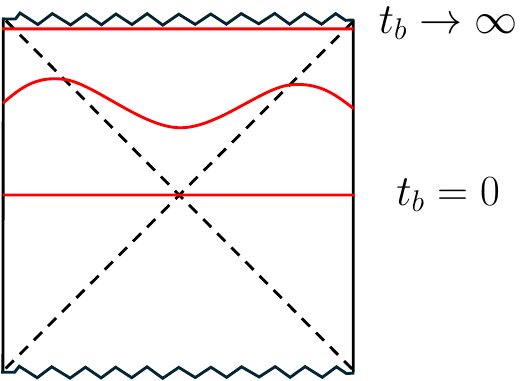}
        \caption{}
        \label{fig:3a}
    \end{subfigure}
    \hspace{0.09\textwidth}
    \begin{subfigure}[b]{0.4\textwidth}
        \includegraphics[width=\textwidth]{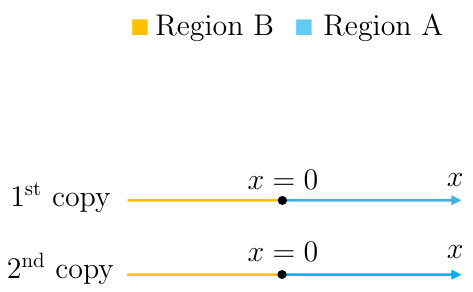}
        \caption{}
        \label{fig:3b}
    \end{subfigure}
    \caption{(a) Extremal curves in BTZ (b) Schematic visualization showing the spatial directions of the two copies of the boundary CFT. Each copy is divided into two halves at $x = 0$. We are interested in the entanglement between $A$ and $B$.} 
    \label{fig:sidebyside}
\end{figure}

The Penrose diagram of an eternal non-rotating BTZ black hole 
 has two boundaries (see {\it e.g.}, Fig. \ref{fig:3a}) --- representing two entangled copies of the boundary CFT$_2$ in a thermofield double state
\begin{equation}
    |\Psi\rangle=\sum_n\left|E_n\right\rangle_1\left|E_n\right\rangle_2 e^{-\frac{\beta}{2} E_n}~,
    \label{Eq:TotState}
\end{equation}
where 1 and 2 label the two copies of the boundary CFT$_2$ and $\beta$ is the inverse temperature (periodicity of the Euclidean time). The corresponding density matrix is given by $\rho_{total
} = |\Psi\rangle \langle\Psi|$. One can then divide the whole system into different regions and study entanglement between them. A convenient way to partition the system is as follows. At a fixed boundary time $t_b$, each boundary --- parametrized by the coordinate $x$ --- is divided into two halves at $x=0$. The regions with $x>0$ on both boundaries are labeled as region $A$, while the regions with $x<0$ are labeled as region $B$ (see {\it e.g.}, Fig. \ref{fig:3b}). The entanglement between regions $A$ and $B$ is then given by
$S_A = -\operatorname{Tr} \rho_A \log \rho_A$~,
where $\rho_A = \operatorname{Tr}_B \rho_{\text{total}}$ is the reduced density matrix of subsystem $A$. It's easy to see that $S_A = S_B$. In the discussion that follows, we will set up a holographic computation of the time evolution of the entanglement entropy between regions A and B in a quenched system and focus mainly on its late-time behavior.

According to the Hubeny-Rangamani-Takayanagi (HRT) proposal \cite{Ryu:2006bv, Hubeny:2007xt}, the entanglement entropy between regions A and B is computed by extremizing the area functional 
\begin{equation}\label{See}
    S_e = \frac{1}{4G_N^{(10)}} \int_\Sigma d^8\sigma e^{-(\Phi-\Phi_0)}\sqrt{G}~,
\end{equation}
where $G_N^{(10)}$ is the 10-dimensional Newton constant, $\Sigma$ is a codimension-2 (in this case $\Sigma$ is  8-dimensional) hypersurface wrapping the internal space $S^3\times T^4$ and anchored at the two boundaries of eternal BTZ at some boundary time $t_b$ (the red curves in Fig. \ref{fig:3a}), and $G$ is the determinant of the induced (string-frame) metric on that hypersurface. At large boundary time, the entanglement entropy is expected to grow linearly in $t_b$ \cite{Hartman:2013qma}: 
\begin{equation}\label{sentg}
    S_e^{BTZ}=\frac{v_e}{v_s} s\,t_b~,
\end{equation}
 where $s$ is the Bekenstein-Hawking entropy density of the BTZ black hole, $v_e$ is the velocity of entanglement, and $v_s$ is the velocity of sound in a CFT$_2$, which, according to the standard normalization, can be set to unity.

We are interested in the evolution of entanglement entropy in time. For this, one takes time to run forward on both copies of the boundary CFT$_2$ and replace $e^{-\frac{\beta}{2} E_n} \rightarrow$ $e^{-\frac{\beta}{2} E_n-2 i E_n t_b}$ in \eqref{Eq:TotState}. On the bulk side, the computation boils down to 
computing the length of an extremal curve $\rho=\rho(t)$ that ends on the two boundaries at $x = 0$ at some boundary time $t_b$ (see Fig. \ref{fig:3a}). The area functional \eqref{eq:BTZspacetime} for $\rho(t)$ takes the form
\begin{equation}
    S_e^{BTZ}= \frac{1}{4G^{(3)}_N} \int d t \sqrt{-M\sinh^2{\rho}+\ell^2\dot{\rho}^2}~,
\label{Eq:ExtrCurveBTZ}
\end{equation}
where $G^{(3)}_N$ is the three-dimensional Newton constant in $AdS_3$ given by
\begin{equation}
    G^{(3)}_N= \frac{G^{(10)}_N}{V_{T^4}V_{S^3}}=\frac{3\ell}{2c}~.
\end{equation}

The extremal curve $\rho(t)$ (the red curves in Fig. \ref{fig:3a}) starts from the right boundary, enters the interior of the horizon, exits the horizon on the other side, and ends at the left boundary. The coordinate patch $(t,\rho,x)$ in \eqref{eq:BTZspacetime}  describes the region outside the horizon on the right wedge.  The interior of the horizon is described by the analytic continuation \eqref{analytic}. Moreover, inside the horizon, the extremal curve has a reflection symmetry under $t_{interior} \to -t_{interior}$, implying  $\dot{\rho}=0$ at $t_{interior}=0$. We denote this interior point by $ \kappa = \kappa_0$.

\begin{figure}[htbp]
    \centering
    \hspace*{-0.2\textwidth} 
\includegraphics[width=0.63\textwidth]{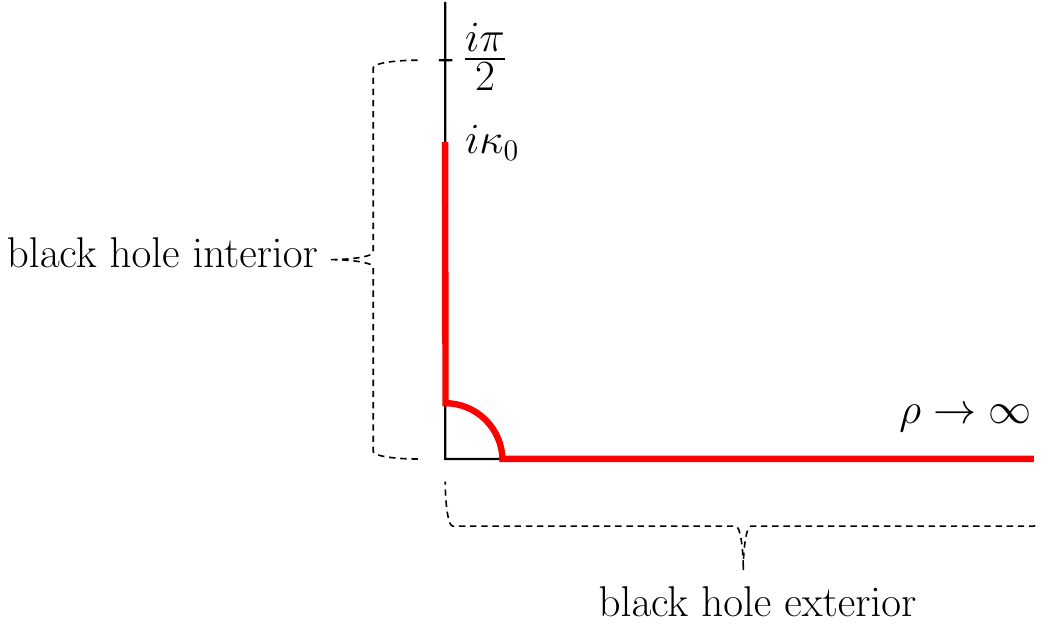}
    \caption{The red curve is the contour of   the $\rho$ integral in the complex $\rho$ plane. The integrand \eqref{tb} has a pole at $\rho=0$, which is avoided by the arc around the origin.
    } 
    \label{fig:rhoComplexPlane}
\end{figure}

Since the action \eqref{Eq:ExtrCurveBTZ} is time independent, the associated Hamiltonian is conserved:
\begin{equation}\label{eom}
    H^{BTZ} = \frac{M\sinh^2{\rho}}{\sqrt{-M\sinh^2{\rho}+\dot{\rho}^2}}=  M^{\frac{1}{2}}\sin{\kappa_0}~,
\end{equation}
where we have used the fact that $\dot{\rho}=0$ at $\rho=i\kappa_0$.
The solution to this equation is given by
\begin{equation}\label{tb}
    t_b= -\frac{i}{4 T}-\int_{i \kappa_0}^\infty \frac{\ell d \rho}{M^{\frac{1}{2}}\sinh\rho \sqrt{1+\frac{\sinh^2\rho}{\sin^2{\kappa_0}}}} = \frac{\ell}{M^{\frac{1}{2}}}\tanh ^{-1}(\sin \kappa_0)~,
\end{equation}
where the $\rho$ integral runs over the contour shown in Fig. \ref{fig:rhoComplexPlane}. Note that the contour integral has a pole at $\rho=0$ and must be avoided as shown in Fig. \ref{fig:rhoComplexPlane}.

There is a nice, intuitive trick discussed in \cite{Hartman:2013qma} for evaluating the extremum of the integral \eqref{Eq:ExtrCurveBTZ} at late times. 
As time progresses, more and more of the curve sits in the interior region of the black hole. 
At a very late boundary time $t_b$\textcolor{red}{}{,} the curve sits at $\kappa \sim \kappa_m$ for a long period of the spacelike $t_{\mathrm{interior}}$ and then crosses the horizon and goes all the way to the two boundaries. 
The portion of the extremal curve that crosses the horizon and goes all the way to the boundary is independent of the boundary time $t_b$, and we discard it\footnote{This portion of the curve is, however, divergent, which corresponds to the standard short-distance divergence (coming from the $\rho \to \infty$ limit of the area functional \eqref{Eq:ExtrCurveBTZ}) of entanglement entropy. Since we are only interested in time-dependent terms in the entanglement entropy, we subtract this term from our result.}.  
The interior segment of the curve that sits at a fixed $\kappa = \kappa_m$ has $\dot{\rho} = 0$ (because $\rho \equiv i\kappa$ stops changing and $\kappa_m$ is the maximum $\kappa_0$). 
This means its length grows linearly in $t_b$. 
Indeed, at large $t_b$, the time-dependent contribution to the entanglement entropy obtained by setting $\dot{\rho} = 0$ and $\rho = i\kappa_m$ in \eqref{Eq:ExtrCurveBTZ} takes the form
\begin{equation}\label{seebtz}
    S_e^{BTZ} = \frac{ M^{\frac{1}{2}} \sin(\kappa_m)t_b}{4G_N^{(3)}} = \frac{\pi c\,  T \sin(\kappa_m) t_b}{3} +O(t_b^0)~. 
\end{equation}
Since $\lim_{\kappa_0\to \kappa_m}t_b(\kappa_0)\to \infty$, it follows from \eqref{tb} that 
\begin{equation}\label{kmbtz}
    \kappa_m=\frac{\pi}{2}~,
\end{equation}
which can easily be recognized as the BTZ singularity from the geometry \eqref{eq:BTZspacetime} with the analytic continuation \eqref{analytic}.\footnote{The late time extremal curve in BTZ hugs the singularity. This feature is typical of BTZ black holes and is no longer true for AdS black holes in higher dimensions, where $\kappa_m$ lies some finite distance away from the singularity. BTZ black holes are special because the curvature is constant everywhere except at the singularity, where it diverges.} Substituting \eqref{kmbtz} in \eqref{seebtz} one obtains the late time behavior of entanglement entropy 
\begin{equation}\label{seebtz2}
    S_e^{BTZ}= \frac{\pi c  T}{3}t_b  = \frac{   S^{BTZ}}{2\pi R}t_b  = s\, t_b ~,
\end{equation}
where  $s = \frac{S^{BTZ}}{2\pi R}$ is the Bekenstein-Hawking entropy density of the BTZ black hole. Comparing \eqref{seebtz2} with \eqref{sentg}, one recovers the well-known result $v_e/v_s = 1$ in a CFT$_2$. The late time behavior of entanglement entropy in a thermal CFT$_2$ \eqref{seebtz2} can also be reproduced by an independent CFT calculation \cite{Calabrese:2005in}.

\subsection{Extremal curves in $\mathcal{M}_3^T$}\label{sec3.2}

In this subsection, we will extend the analysis of the previous subsection \ref{sec3.1}, and calculate the late time extremal curve extending from one boundary to the other in the  spacetime $\mathcal{M}_3^T$ given by \eqref{eq:RELEVMetrM3T}
\begin{equation}
\begin{aligned}
    &d s^2_{\mathcal{M}}=-  g(\rho)^2d t^2+h(\rho)^2 d x^2+k \alpha^{\prime} d \rho^2~,\quad e^{-2(\Phi-\Phi_0)} = j(\rho)~,\\
    &g(\rho)^2 = \frac{U_T^2 \sinh ^2 \rho}{j(\rho)},\quad h(\rho)^2 = \frac{U_T^2 \cosh ^2 \rho}{j(\rho)}, \quad j(\rho) = 1+U_T^2 \cosh ^2 \rho~,
    & 
\end{aligned}    
\end{equation}
 and argue why it can be interpreted as the time-evolved entanglement entropy of quenched boundary field theory (LST). Before we begin our calculation, let us remind the reader of certain facts regarding the Ryu-Takayanagi (RT) proposal for non-conformal field theories and non-AdS spacetime, in particular, asymptotically linear-dilaton spacetime. 

It was argued in \cite{Ryu:2006ef,Klebanov:2007ws} that the RT proposal computes the entanglement entropy of a certain class of non-conformal field theory dual to classical string theory in a spacetime that doesn't asymptote to AdS reasonably well. Moreover, the RT curve has been calculated in the spacetime \eqref{eq:RELEVMetrM3T} \cite{Chakraborty:2018kpr,Asrat:2020uib} and shown to agree perturbatively with a boundary field theory calculation. The RT curve in a certain non-Lorentz invariant generalization of asymptotically linear dilaton spacetime has also been investigated \cite{Asrat:2019end,Chakraborty:2020udr} and has been shown to agree perturbatively with the boundary field theory expectation. That being said, it is natural to expect that the extremal curve in $\mathcal{M}_3^T$, connecting the two boundaries at a fixed boundary time $t_b$, captures the time-dependent entanglement entropy in the quenched boundary field theory. However, this proposal rests on somewhat less firm ground, as—unlike previous studies—we currently lack an independent field theory computation to substantiate it. In the discussion that follows, we will provide the derivation of the late-time behavior of extremal curves in $\mathcal{M}_3^T$ and discuss its implications from the perspective of the boundary LST. 

The extremal curve connecting the two boundaries in $\mathcal{M}_3^T$ at some boundary time $t_b$ is obtained by extremizing the area functional \eqref{See}
\begin{align}
    S_e=\frac{c}{6 \sqrt{k\alpha'}}\int d t j(\rho) \sqrt{-g(\rho)^2 +\ell^2 \dot{\rho}(t)^2}~.
    \label{Sm3}
\end{align}
Its equation of motion is given by
\begin{equation}
    \frac{g^2 j}{\sqrt{-g^2+k \alpha^{\prime} \dot{\rho}(t)^2}}= - i g_0 j_0~,
\end{equation}
where $g_0=g(\kappa_0),j_0=j(\kappa_0)$ and as before $\dot{\rho}=0$ at $\rho=i\kappa_0$. 
As usual, the boundary time, $t_b$, is a function of $\kappa_0$ given by
\begin{align}\label{tbm3}
    t_b(\kappa_0) \equiv t(\infty)=-i \frac{\pi \ell}{2 } \sqrt{\frac{1+U_T^2}{U_T^2}} -\int_{i \kappa_0}^\infty \frac{\ell d \rho
    }{g \sqrt{1-\frac{g^2 j^2}{g_0^2 j_0^2}}}~,
\end{align}
where the contour for the $\rho$ integral is given the thick red curve in Fig.  \ref{fig:rhoComplexPlane}.

Following the same line of argument as in the case of BTZ at late boundary time, the length of the extremal curve is obtained by setting $\dot{\rho}=0$ in \eqref{Sm3} and evaluating the integrand at $\rho=i\kappa_m$ (see the discussion above \eqref{seebtz})
\begin{equation}\label{seee}
   S_e= \frac{c}{6\ell}\,a(\kappa_m)t_b~,
\end{equation}
where
\begin{equation}
    a(\kappa_m) = \frac{T\sin \kappa_m\sqrt{T_H^2-T^2\sin^2{\kappa_m}}}{T_H^2-T^2}~,
   \label{Eq:areaGENERAL}
\end{equation}
and $\kappa_m$ is obtained by extremizing $a(\kappa_0)$:
\begin{equation}\label{kappam}
    \kappa_m^{(1)} = \frac{\pi}{2}, \qquad \kappa_m^{(2)} =\sin^{-1}\left(\frac{T_c}{T}\right)~,
\end{equation}
where 
\begin{equation}
    T_c=\frac{T_H}{\sqrt{2}}~.
\end{equation}
Interestingly, unlike BTZ, here we obtain two different solutions. 
Naively, one might then think that there are two real extremal curves that connect the two boundaries; however, a closer inspection of the second solution, $\kappa_m^{(2)}$, in \eqref{kappam} indicates that a real solution may only exist for $T\geq T_c$. Moreover, it's not guaranteed that the curve associated with the first solution $\kappa_m^{(1)}$ is real for all $T<T_H$. Therefore, we should carefully investigate the features of the extremal curves in the complex $t_{b}$ and $\kappa_0$ plane as a function of temperature.

To identify $t_b$ associated with each curve, we consider large values of $t_b$ for which the curve's reflection symmetric point approaches $\kappa_0 \to \kappa_m$. Choosing $\kappa_m^{(1)}= \frac{\pi}{2}$ allows us to investigate the first curve, while taking $
\kappa_m^{(2)} = \sin^{-1}\left( \frac{T_c}{T} \right)$
corresponds to the second curve. However, due to divergences in the integrand near $\kappa_m$, it is appropriate instead to  set
$\kappa_0 = \kappa_m^{(1,2)} - \epsilon$
for small, real, and positive $\epsilon$ and study the $t_{b}$ as a function of $T$ in the complex $t_b$ plane. 

Varying the temperature in the range $0 < T < T_H$, the numerical evaluation of \eqref{tbm3}
reveals a surprising result: the curves are not always real. We separately discuss the fate of the two curves (associated with $\kappa_{m}^{(1,2)}$) as follows.

\begin{figure}[t]
    \centering
    \includegraphics[width=0.53\textwidth]{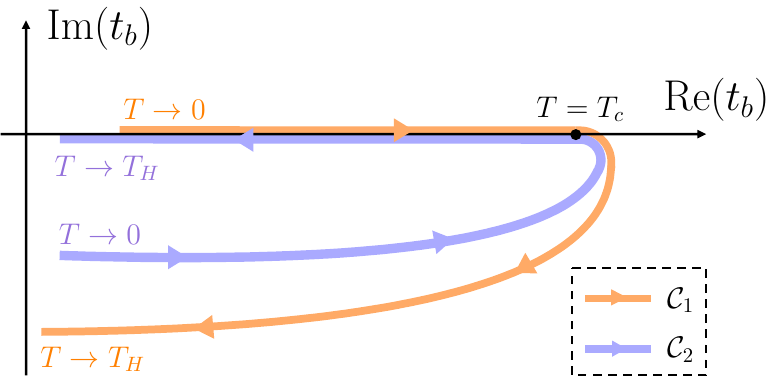}
    \caption{Complex boundary time $t_b$ parametrized by temperature $T$. The orange line corresponds to the curve $\mathcal{C}_1$ (with $\kappa_m = \frac{\pi}{2}$), where $t_b$ is real for $T \leq T_c$ and becomes complex for $T > T_c$. The purple line represents the curve $\mathcal{C}_2$   (with $\kappa_m = \sin^{-1}(T_c/T)$), where $t_b$ is complex for $  T< T_c$ and becomes real for $T_c\leq T < T_H$. The two curves intersect at the critical temperature $T = T_c$, where both curves are real. Arrows indicate the direction of increasing temperature. 
    } 
    \label{fig:ComplexPlane}
\end{figure}

\begin{itemize}
    \item \underline{ $\kappa_m^{(1)}$-curve ($\mathcal{C}_1$)}: As temperature increases, this curve initially yields real values of $t_b$. However, this reality persists only up to a critical temperature $T = T_c$. Beyond this point, $t_b$ becomes complex and remains complex up to the maximal temperature $T = T_H$ (see the orange curve in  Fig.~\ref{fig:ComplexPlane}). This signals that the length of the curve  $\mathcal{C}_1$ can no longer be interpreted as a measure of entanglement entropy at high temperature. For $T\leq T_c$, however, the length of this curve accurately captures the time-dependent entanglement entropy. At low temperature ({\it i.e.,} $T\ll T_c$), it reproduced the BTZ result or, more appropriately, the time-dependent entanglement entropy of the boundary CFT$_2$.
    
    \item \underline{$\kappa_m^{(2)}$-curve ($\mathcal{C}_2$)}: In contrast, this curve initially yields complex values of $t_b$ for $T < T_c$. As the temperature increases past the critical point $T = T_c$, the values of $t_b$ become real and remain real for all $T_c\leq T <  T_H$ (see the purple curve in Fig.~\ref{fig:ComplexPlane}). The reality of $\mathcal{C}_2$ in the regime $T_c\leq T<T_H$ 
    motivates us to propose that its length is likely to capture the physics of entanglement entropy at short distances in the spacetime theory.
    
\end{itemize}

The discussion above indicates that the two curves are never simultaneously
real (except $T=T_c$); rather, they are real in the complementary domains. It is worth stressing that the value of $t_b$ at $T=T_c$ is real and equal for both curves.  Since we expect the entanglement entropy to be real, it is natural to restrict ourselves to the real parts of these curves and smoothly connect them at $T = T_c$. As we will see below, the resulting physics is remarkably rich and may signal the presence of a phase transition as we interpolate between the two solutions. In the discussion that follows, we will adapt this path and restrict ourselves to the real part of the extremal curves.

From \eqref{seee} one can conclude that the late time behavior of the  entanglement takes the form 
\begin{equation}\label{seee1}
   S_e= \frac{c}{6\ell}a(\kappa_m)t_b~,
\end{equation}
where 
\begin{equation}\label{am}
    a(\kappa_m) = \left\{\begin{array}{lll}
      \frac{T}{\sqrt{T_H^2-T^2}}~, & \text { for } &  T\leq T_c~, \\[1em]
    \frac{T_c^2}{T_H^2-T^2}~, & \text { for } & T_c\leq T<T_H~.
\end{array}\right.
\end{equation} 
It is easy to check that $a(\kappa_m)=1$ at $T=T_c$ for both solutions. Substituting \eqref{am} in \eqref{seee1} one obtains
\begin{equation}\label{seee2}
   S_e = \left\{\begin{array}{lll}
      s\,t_b~, & \text { for } &  T\leq T_c~, \\[.5em]
\frac{T_H}{2T}\left(1-\frac{T^2}{T_H^2}\right)^{-\frac{1}{2}} s t_b~, & \text { for } & T_c\leq T<T_H~.
\end{array}\right.
\end{equation} 
where $s$ is the Bekenstein-Hawking entropy density $S/(2\pi R)$ with $S$ given by \eqref{sbek}. It is easy to see that for $T\leq T_c$ the large time behavior of entanglement is exactly as one would expect for a local field theory \eqref{sentg},\eqref{seebtz2} with $v_e/v_s=1$. However, for $T_c\leq T<T_H$ the coefficient of $st_b$ in \eqref{seee2} diverges as $T\to T_H$. This divergence is reminiscent of the fact that the velocity of 
sound in LST \eqref{eq:velOFwave} goes to zero as $\sqrt{T_H^2-T^2}/T_H$ \eqref{eq:velOFwave} as $T$ approaches $T_H$.

\begin{figure}[t]
    \centering
    \begin{subfigure}[b]{0.45\textwidth}
\includegraphics[width=\textwidth]{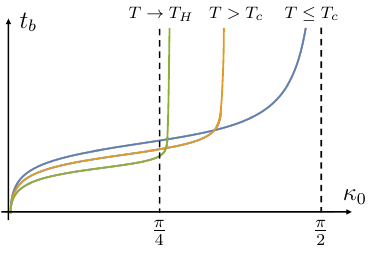}
        \caption{}
        \label{fig:5a}
    \end{subfigure}
    \hspace{0.09\textwidth}
    \begin{subfigure}[b]{0.25\textwidth}
        \includegraphics[width=\textwidth]{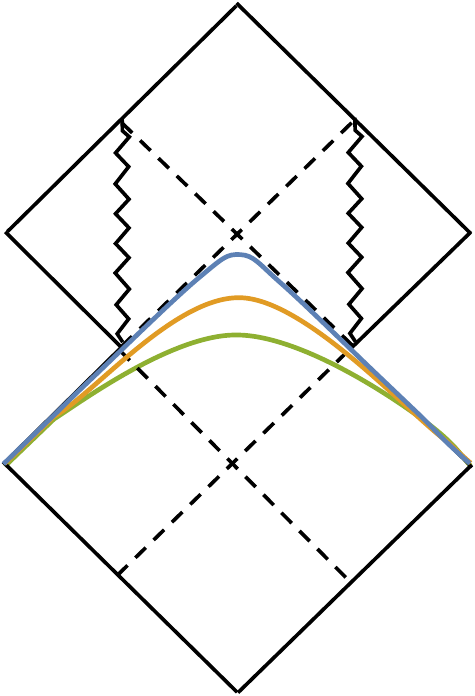}
        \caption{}
        \label{fig:5b}
    \end{subfigure}
    \caption{(a) $t_b$ vs $\kappa_0$ plot (b)  Schematic visualization of the real extremal curves in $\mathcal{M}_3^T$.}
    \label{fig:sidebyside}
\end{figure}

\begin{figure}[t]
    \begin{center}
    \begin{subfigure}[b]{0.328\textwidth}
        \includegraphics[width=\textwidth]{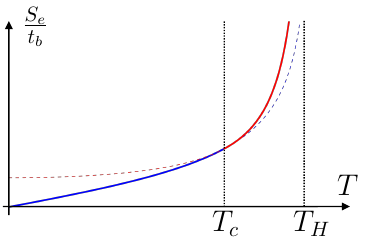}
        \caption{}
        \label{fig:7a}
    \end{subfigure}
    \hfill
    \begin{subfigure}[b]{0.328\textwidth}
        \includegraphics[width=\textwidth]{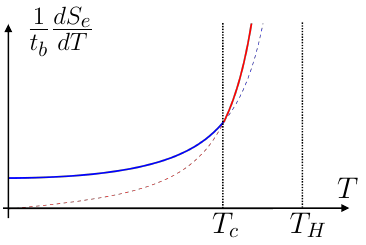}
        \caption{}
        \label{fig:7b}
    \end{subfigure}
    \hfill
    \begin{subfigure}[b]{0.328\textwidth}
    \includegraphics[width=\textwidth]{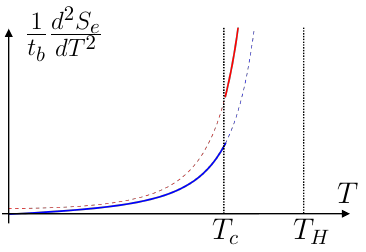}
        \caption{}
        \label{fig:7c}
    \end{subfigure}
    \caption{ Second order phase transition.  }
    \label{fig:PhaseTrans}
    \end{center}
\end{figure}

Following the discussion above, a few comments are in order.
\begin{itemize}
    \item At low temperature ({\it i.e.}, $T\leq T_c$), the curve $\mathcal{C}_1$ is real and at late time it hugs the inner horizon (see the blue curve in Fig. \ref{fig:5a}~\&~\ref{fig:5b}). A similar behavior is observed in BTZ black holes, where this feature is commonly referred to as the `absence of an entanglement shadow.' One can interpret this phenomenon as follows. At low temperature, one can approximate the boundary field theory as a local field theory in two dimensions, and that there is no entanglement shadow is reminiscent of the fact that the boundary field theory is an honest two-dimensional field theory.

    \item At late time, as the temperature goes past $T_c$, the curve $\mathcal{C}_2$ is real. At $T=T_c$, it still hugs the singularity, and as $T$ increases, it starts to leave the inner horizon, creating an entanglement shadow (see the yellow curve in Fig. \ref{fig:5a}~\&~\ref{fig:5b}). As the temperature approached the maximal temperature $T_H$, the curve $\mathcal{C}_2$ saturates at $\kappa_{m}^{(2)}=\pi/4$ (see the green curve in Fig. \ref{fig:5a}~\&~\ref{fig:5b}). Note that entanglement shadow is a feature of higher-dimensional field theories. In our case, the boundary field theory at high temperature ({\it i.e.}, $T_c\leq T< T_H$) is well approximated by a two-dimensional compactification of six-dimensional LST compactified on $T^4$. The entanglement shadow that we observe here is reminiscent of the fact that the spacetime theory is effectively six-dimensional.   
    
\item It is easy to see from \eqref{seee2} that at late time entanglement is continuous at $T=T_c$, and so is its first temperature derivative at a fixed but large $t_b$ (see Fig. \ref{fig:7a} \& \ref{fig:7b}). However, its second derivative is discontinuous at $T=T_c$ (see Fig. \ref{fig:7c}). It is tempting to refer to this discontinuity as a second-order phase transition separating the local phase (for $T\leq T_c$) from the non-local phase (for $T_c\leq T<T_H$). However, a microscopic understanding of this phase transition is beyond the scope of this paper. A similar second-order phase transition has also been reported in the analysis of the expectation value of the holographic Wilson-loop operator in $\mathcal{M}_3^T$ \cite{Chakraborty:2018aji}, where one needs to extremize the worldvolume of a D1 brane hanging in the bulk with its end-points anchored at the boundary.  

\item It may be alarming to the reader that a SUGRA background \eqref{eq:RELEVMetrM3T}, which is smooth everywhere outside the inner horizon and analytic for $0\leq T <T_H$, gives rise to non-analyticity of extremal curves at $T=T_c <T_H$. However, a closer inspection reveals that both curves $\mathcal{C}_{1,2}$ are individually analytic in the complex $(t,\rho)$ space for $0\leq T <T_H$. The non-analyticity discussed in the item above appears when we restrict the discussion to real extremal curves, where extremal solutions $\mathcal{C}_{1,2}$ are real in the complementary temperature domain.
    
\end{itemize}

\section{Discussion}\label{diss}

In this paper, we propose that the length of real extremal curves connecting the two boundaries of an eternal black hole at a fixed boundary time in single-trace $T\bar{T}$-holography encodes the time-evolved entanglement entropy of an entangled, quenched boundary system.   
We studied extremal curves connecting the two boundaries at late times and found that there are two such solutions in the maximally extended complexified geometry. 
Both of these solutions are analytic in temperature in the complexified geometry.  
When restricted to the real geometry, we found that the extremal curves are real in complementary temperature domains, which induces non-analyticity at the critical temperature $T = T_c$. 
We interpret this non-analyticity as a second-order phase transition separating the local phase (CFT$_2$-phase) from the non-local phase (LST-phase). 
We have not provided a microscopic understanding of this phase transition. 
Needless to say, it would be good to have an independent microscopic understanding of this phenomenon.

In the standard AdS/CFT, Lewkowycz-Maldacena \cite{Lewkowycz:2013nqa} proved the Ryu-Takayanagi proposal by continuing the replica trick in the bulk path integral and then taking the R\'enyi index $n\to 1$. In particular, they showed that there is a smooth $n\to 1$ limit that reproduces the standard Ryu-Takayanagi proposal for entanglement entropy. The existence of two extremal curves in the spacetime $\mathcal{M}_3^T$---each real in complementary temperature regimes---raises important questions about the smoothness of the $n \to 1$ limit in the holographic replica trick for a spacetime that doesn't asymptote to AdS. It would be good to understand the holographic replica trick in the spacetime geometry under consideration here.

While this paper offers a detailed investigation of the late-time behavior of extremal curves in single-trace $T\bar{T}$-holography, the early-time behavior remains unexplored. The reason for this omission is primarily conceptual rather than technical. In a quenched CFT, it is meaningful to discuss the early-time behavior of observables \cite{Liu:2013iza}—specifically, timescales shorter than the equilibrium scale. However, in LST, which is inherently non-local and characterized by its own non-locality scale, the notion of ``early time" becomes less clear. The interplay between the non-locality scale and the equilibrium scale introduces additional subtleties, making the analysis more intricate. It would be nice to have a good understanding of this issue.   

There are non-Lorentz invariant variants of single-trace $T\bar{T}$-holography \cite{Chakraborty:2019mdf}. Black holes in such setups have been constructed in \cite{Apolo:2021wcn}. It would be interesting to extend the analysis in this paper to non-Lorentz invariant setups and investigate how the lack of Lorentz invariance affects the late-time behavior of extremal curves.

Recently, there has been renewed interest in exploring holography in more general backgrounds with null boundaries, particularly in the context of non-commutative open string theory in the decoupling limit \cite{Georgescu:2024iam,Ferko:2025elh}. It would be interesting to analyze extremal curves, in particular the standard RT curve, in these contexts.

\acknowledgments
We thank A. Hashimoto, D. Kutasov, H. Liu, and E. Martinec for insightful discussions. We are especially grateful to J. Maldacena for useful comments. The work of SC and GP is supported by the Department of Physics at The Ohio State University. MM is funded by the presidential
fellowship granted by the Ohio State University.



\begin{thebibliography}{10}

\bibitem{Giveon:2017nie}
A.~Giveon, N.~Itzhaki and D.~Kutasov, \emph{{$ \mathrm{T}\overline{\mathrm{T}}
  $ and LST}}, \href{https://doi.org/10.1007/JHEP07(2017)122}{\emph{JHEP}
  {\bfseries 07} (2017) 122}
  [\href{https://arxiv.org/abs/1701.05576}{{\ttfamily 1701.05576}}].

\bibitem{Kutasov:1999xu}
D.~Kutasov and N.~Seiberg, \emph{{More comments on string theory on AdS(3)}},
  \href{https://doi.org/10.1088/1126-6708/1999/04/008}{\emph{JHEP} {\bfseries
  04} (1999) 008} [\href{https://arxiv.org/abs/hep-th/9903219}{{\ttfamily
  hep-th/9903219}}].

\bibitem{Chakraborty:2024mls}
S.~Chakraborty, A.~Giveon and A.~Hashimoto, \emph{{Thermal partition function
  of $ {J}_3{\overline{J}}_3 $ deformed AdS$_{3}$}},
  \href{https://doi.org/10.1007/JHEP07(2024)141}{\emph{JHEP} {\bfseries 07}
  (2024) 141} [\href{https://arxiv.org/abs/2403.03979}{{\ttfamily
  2403.03979}}].

\bibitem{Apolo:2019zai}
L.~Apolo, S.~Detournay and W.~Song, \emph{{TsT, $T\bar{T}$ and black strings}},
  \href{https://doi.org/10.1007/JHEP06(2020)109}{\emph{JHEP} {\bfseries 06}
  (2020) 109} [\href{https://arxiv.org/abs/1911.12359}{{\ttfamily
  1911.12359}}].

\bibitem{Guica:2019nzm}
M.~Guica and R.~Monten, \emph{{$T\bar T$ and the mirage of a bulk cutoff}},
  \href{https://doi.org/10.21468/SciPostPhys.10.2.024}{\emph{SciPost Phys.}
  {\bfseries 10} (2021) 024}
  [\href{https://arxiv.org/abs/1906.11251}{{\ttfamily 1906.11251}}].

\bibitem{Jiang:2019epa}
Y.~Jiang, \emph{{A pedagogical review on solvable irrelevant deformations of 2D
  quantum field theory}},
  \href{https://doi.org/10.1088/1572-9494/abe4c9}{\emph{Commun. Theor. Phys.}
  {\bfseries 73} (2021) 057201}
  [\href{https://arxiv.org/abs/1904.13376}{{\ttfamily 1904.13376}}].

\bibitem{He:2025ppz}
S.~He, Y.~Li, H.~Ouyang and Y.~Sun, \emph{{$T\overline{T}$ Deformation:
  Introduction and Some Recent Advances}},
  \href{https://arxiv.org/abs/2503.09997}{{\ttfamily 2503.09997}}.

\bibitem{Chakraborty:2018kpr}
S.~Chakraborty, A.~Giveon, N.~Itzhaki and D.~Kutasov, \emph{{Entanglement
  beyond AdS}},
  \href{https://doi.org/10.1016/j.nuclphysb.2018.08.011}{\emph{Nucl. Phys. B}
  {\bfseries 935} (2018) 290}
  [\href{https://arxiv.org/abs/1805.06286}{{\ttfamily 1805.06286}}].

\bibitem{Asrat:2020uib}
M.~Asrat and J.~Kudler-Flam, \emph{{$T\bar{T}$, the entanglement wedge cross
  section, and the breakdown of the split property}},
  \href{https://doi.org/10.1103/PhysRevD.102.045009}{\emph{Phys. Rev. D}
  {\bfseries 102} (2020) 045009}
  [\href{https://arxiv.org/abs/2005.08972}{{\ttfamily 2005.08972}}].

\bibitem{Asrat:2019end}
M.~Asrat, \emph{{Entropic $c${\textendash}functions in $T{\bar T}, J{\bar T},
  T{\bar J}$ deformations}},
  \href{https://doi.org/10.1016/j.nuclphysb.2020.115186}{\emph{Nucl. Phys. B}
  {\bfseries 960} (2020) 115186}
  [\href{https://arxiv.org/abs/1911.04618}{{\ttfamily 1911.04618}}].

\bibitem{Chakraborty:2020udr}
S.~Chakraborty and A.~Hashimoto, \emph{{Entanglement entropy for $
  \mathrm{T}\overline{\mathrm{T}} $, $ \mathrm{J}\overline{\mathrm{T}} $, $
  \mathrm{T}\overline{\mathrm{J}} $ deformed holographic CFT}},
  \href{https://doi.org/10.1007/JHEP02(2021)096}{\emph{JHEP} {\bfseries 02}
  (2021) 096} [\href{https://arxiv.org/abs/2010.15759}{{\ttfamily
  2010.15759}}].

\bibitem{Hartman:2013qma}
T.~Hartman and J.~Maldacena, \emph{{Time Evolution of Entanglement Entropy from
  Black Hole Interiors}},
  \href{https://doi.org/10.1007/JHEP05(2013)014}{\emph{JHEP} {\bfseries 05}
  (2013) 014} [\href{https://arxiv.org/abs/1303.1080}{{\ttfamily 1303.1080}}].

\bibitem{Maldacena:1996ky}
J.M.~Maldacena, \emph{{Black holes in string theory}}, Ph.D. thesis, Princeton
  U., 1996.
\newblock \href{https://arxiv.org/abs/hep-th/9607235}{{\ttfamily
  hep-th/9607235}}.

\bibitem{Hyun:1997jv}
S.~Hyun, \emph{{U duality between three-dimensional and higher dimensional
  black holes}}, {\emph{J. Korean Phys. Soc.} {\bfseries 33} (1998) S532}
  [\href{https://arxiv.org/abs/hep-th/9704005}{{\ttfamily hep-th/9704005}}].

\bibitem{Giveon:1998ns}
A.~Giveon, D.~Kutasov and N.~Seiberg, \emph{{Comments on string theory on
  AdS(3)}}, \href{https://doi.org/10.4310/ATMP.1998.v2.n4.a3}{\emph{Adv. Theor.
  Math. Phys.} {\bfseries 2} (1998) 733}
  [\href{https://arxiv.org/abs/hep-th/9806194}{{\ttfamily hep-th/9806194}}].

\bibitem{Chakraborty:2020swe}
S.~Chakraborty, A.~Giveon and D.~Kutasov, \emph{{$ T\overline{T} $, black holes
  and negative strings}},
  \href{https://doi.org/10.1007/JHEP09(2020)057}{\emph{JHEP} {\bfseries 09}
  (2020) 057} [\href{https://arxiv.org/abs/2006.13249}{{\ttfamily
  2006.13249}}].

\bibitem{Chakraborty:2023mzc}
S.~Chakraborty, A.~Giveon and D.~Kutasov, \emph{{Comments on single-trace $
  T\overline{T} $ holography}},
  \href{https://doi.org/10.1007/JHEP06(2023)018}{\emph{JHEP} {\bfseries 06}
  (2023) 018} [\href{https://arxiv.org/abs/2303.12422}{{\ttfamily
  2303.12422}}].

\bibitem{Chakraborty:2023zdd}
S.~Chakraborty, A.~Giveon and D.~Kutasov, \emph{{Momentum in Single-trace
  $T\bar T$ Holography}},
  \href{https://doi.org/10.1016/j.nuclphysb.2023.116405}{\emph{Nucl. Phys. B}
  {\bfseries 998} (2024) 116405}
  [\href{https://arxiv.org/abs/2304.09212}{{\ttfamily 2304.09212}}].

\bibitem{Chakraborty:2020fpt}
S.~Chakraborty, G.~Katoch and S.R.~Roy, \emph{{Holographic complexity of LST
  and single trace $ T\overline{T} $}},
  \href{https://doi.org/10.1007/JHEP03(2021)275}{\emph{JHEP} {\bfseries 03}
  (2021) 275} [\href{https://arxiv.org/abs/2012.11644}{{\ttfamily
  2012.11644}}].

\bibitem{Parnachev:2005hh}
A.~Parnachev and A.~Starinets, \emph{{The Silence of the little strings}},
  \href{https://doi.org/10.1088/1126-6708/2005/10/027}{\emph{JHEP} {\bfseries
  10} (2005) 027} [\href{https://arxiv.org/abs/hep-th/0506144}{{\ttfamily
  hep-th/0506144}}].

\bibitem{Goykhman:2013oja}
M.~Goykhman and A.~Parnachev, \emph{{Stringy holography at finite density}},
  \href{https://doi.org/10.1016/j.nuclphysb.2013.05.011}{\emph{Nucl. Phys. B}
  {\bfseries 874} (2013) 115}
  [\href{https://arxiv.org/abs/1304.4496}{{\ttfamily 1304.4496}}].

\bibitem{Chakraborty:2022dgm}
S.~Chakraborty and M.~Goykhman, \emph{{Solvable time-like cosets and holography
  beyond AdS}}, \href{https://doi.org/10.1007/JHEP08(2022)244}{\emph{JHEP}
  {\bfseries 08} (2022) 244}
  [\href{https://arxiv.org/abs/2204.03024}{{\ttfamily 2204.03024}}].

\bibitem{Datta:2018thy}
S.~Datta and Y.~Jiang, \emph{{$T\bar{T}$ deformed partition functions}},
  \href{https://doi.org/10.1007/JHEP08(2018)106}{\emph{JHEP} {\bfseries 08}
  (2018) 106} [\href{https://arxiv.org/abs/1806.07426}{{\ttfamily
  1806.07426}}].

\bibitem{Chakraborty:2020xyz}
S.~Chakraborty and A.~Hashimoto, \emph{{Thermodynamics of $
  \mathrm{T}\overline{\mathrm{T}} $, $ \mathrm{J}\overline{\mathrm{T}} $, $
  \mathrm{T}\overline{\mathrm{J}} $ deformed conformal field theories}},
  \href{https://doi.org/10.1007/JHEP07(2020)188}{\emph{JHEP} {\bfseries 07}
  (2020) 188} [\href{https://arxiv.org/abs/2006.10271}{{\ttfamily
  2006.10271}}].

\bibitem{RangamaniTakayanagi2017}
M.~Rangamani and T.~Takayanagi, \emph{Holographic Entanglement Entropy},
  vol.~931 of \emph{Lecture Notes in Physics}, Springer, Cham (2017),
  \href{https://doi.org/10.1007/978-3-319-52573-0}{10.1007/978-3-319-52573-0}.

\bibitem{Calabrese:2005in}
P.~Calabrese and J.L.~Cardy, \emph{{Evolution of entanglement entropy in
  one-dimensional systems}},
  \href{https://doi.org/10.1088/1742-5468/2005/04/P04010}{\emph{J. Stat. Mech.}
  {\bfseries 0504} (2005) P04010}
  [\href{https://arxiv.org/abs/cond-mat/0503393}{{\ttfamily
  cond-mat/0503393}}].

\bibitem{Ryu:2006bv}
S.~Ryu and T.~Takayanagi, \emph{{Holographic derivation of entanglement entropy
  from AdS/CFT}},
  \href{https://doi.org/10.1103/PhysRevLett.96.181602}{\emph{Phys. Rev. Lett.}
  {\bfseries 96} (2006) 181602}
  [\href{https://arxiv.org/abs/hep-th/0603001}{{\ttfamily hep-th/0603001}}].

\bibitem{Hubeny:2007xt}
V.E.~Hubeny, M.~Rangamani and T.~Takayanagi, \emph{{A Covariant holographic
  entanglement entropy proposal}},
  \href{https://doi.org/10.1088/1126-6708/2007/07/062}{\emph{JHEP} {\bfseries
  07} (2007) 062} [\href{https://arxiv.org/abs/0705.0016}{{\ttfamily
  0705.0016}}].

\bibitem{Ryu:2006ef}
S.~Ryu and T.~Takayanagi, \emph{{Aspects of Holographic Entanglement Entropy}},
  \href{https://doi.org/10.1088/1126-6708/2006/08/045}{\emph{JHEP} {\bfseries
  08} (2006) 045} [\href{https://arxiv.org/abs/hep-th/0605073}{{\ttfamily
  hep-th/0605073}}].

\bibitem{Klebanov:2007ws}
I.R.~Klebanov, D.~Kutasov and A.~Murugan, \emph{{Entanglement as a probe of
  confinement}},
  \href{https://doi.org/10.1016/j.nuclphysb.2007.12.017}{\emph{Nucl. Phys. B}
  {\bfseries 796} (2008) 274}
  [\href{https://arxiv.org/abs/0709.2140}{{\ttfamily 0709.2140}}].

\bibitem{Chakraborty:2018aji}
S.~Chakraborty, \emph{{Wilson loop in a $T\bar{T}$ like deformed
  $\rm{CFT}_2$}},
  \href{https://doi.org/10.1016/j.nuclphysb.2018.12.003}{\emph{Nucl. Phys. B}
  {\bfseries 938} (2019) 605}
  [\href{https://arxiv.org/abs/1809.01915}{{\ttfamily 1809.01915}}].

\bibitem{Lewkowycz:2013nqa}
A.~Lewkowycz and J.~Maldacena, \emph{{Generalized gravitational entropy}},
  \href{https://doi.org/10.1007/JHEP08(2013)090}{\emph{JHEP} {\bfseries 08}
  (2013) 090} [\href{https://arxiv.org/abs/1304.4926}{{\ttfamily 1304.4926}}].

\bibitem{Liu:2013iza}
H.~Liu and S.J.~Suh, \emph{{Entanglement Tsunami: Universal Scaling in
  Holographic Thermalization}},
  \href{https://doi.org/10.1103/PhysRevLett.112.011601}{\emph{Phys. Rev. Lett.}
  {\bfseries 112} (2014) 011601}
  [\href{https://arxiv.org/abs/1305.7244}{{\ttfamily 1305.7244}}].

\bibitem{Chakraborty:2019mdf}
S.~Chakraborty, A.~Giveon and D.~Kutasov, \emph{{$T\bar{T}$, $J\bar{T}$,
  $T\bar{J}$ and String Theory}},
  \href{https://doi.org/10.1088/1751-8121/ab3710}{\emph{J. Phys. A} {\bfseries
  52} (2019) 384003} [\href{https://arxiv.org/abs/1905.00051}{{\ttfamily
  1905.00051}}].

\bibitem{Apolo:2021wcn}
L.~Apolo and W.~Song, \emph{{TsT, black holes, and $ T\overline{T} $ + $
  J\overline{T} $ + $ T\overline{J} $}},
  \href{https://doi.org/10.1007/JHEP04(2022)177}{\emph{JHEP} {\bfseries 04}
  (2022) 177} [\href{https://arxiv.org/abs/2111.02243}{{\ttfamily
  2111.02243}}].

\bibitem{Georgescu:2024iam}
S.~Georgescu, M.~Guica and N.~Kovensky, \emph{{Ascending the attractor flow in
  the D1-D5 system}},  \href{https://arxiv.org/abs/2401.01298}{{\ttfamily
  2401.01298}}.

\bibitem{Ferko:2025elh}
C.~Ferko and S.~Sethi, \emph{{Holography with Null Boundaries}},
  \href{https://arxiv.org/abs/2506.20765}{{\ttfamily 2506.20765}}.

\end{thebibliography}

\providecommand{\href}[2]{#2}\begingroup\raggedright\endgroup

\end{document}